\documentclass[final,5p,times,twocolumn]{elsarticle}

\usepackage{amssymb}

\journal{Physics Letters B}

\begin{document}

\begin{frontmatter}

\title{Microscopic origin of shape coexistence in the N=90, Z=64 region}


\author{Dennis Bonatsos[1], K.E. Karakatsanis[1,2,3], Andriana Martinou[1], T.J. Mertzimekis[4] and N. Minkov[5]}

\address[1]{Institute of Nuclear and Particle Physics, National Centre for Scientific Research 
``Demokritos'', GR-15310 Aghia Paraskevi, Attiki, Greece}

\address[2]{Department of Physics, Faculty of Science, University of Zagreb, HR-10000 Zagreb, Croatia}

\address[3]{Physics Department, Aristotle University of Thessaloniki, Thessaloniki GR-54124, Greece}

\address[4]{Department of Physics, National and Kapodistrian University of Athens, Zografou Campus, GR-15784 Athens, Greece}

\address[5]{Institute of Nuclear Research and Nuclear Energy, Bulgarian Academy of Sciences, 72 Tzarigrad Road, 1784 Sofia, Bulgaria}

\begin{abstract}

A microscopic explanation of the nature of shape coexistence in the N=90, Z=64 region is suggested, based on calculations of single particle energies through standard covariant density functional theory. It is suggested that shape coexistence in the N=90 region is caused by the protons, which create neutron particle-hole (p-h) excitations across the N=112 3-dimensional isotropic harmonic oscillator (3D-HO) magic number, signaling the start of the occupation of the 1i13/2 intruder orbital, which triggers stronger proton-neutron interaction, causing the onset of the deformation and resulting in the shape/phase transition from  spherical to deformed nuclei described by the X(5) critical point symmetry. A similar effect is seen  in the N=60, Z=40 region, in which  p-h excitations across the N=70 3D-HO magic number occur, signaling the start of the occupation of the 1h11/2 intruder orbital. 

\end{abstract}


\end{frontmatter}

\section{Introduction}

Shape coexistence \cite{Wood,Heyde,Garrett}, appearing in a nucleus when the ground state band and another band lie very close in energy but possess radically different structures, for example one of them being near-spherical and the other deformed, is a field of intense current experimental investigations \cite{Garrett}, since its occurrence appears to be connected to subtle properties of the nucleon-nucleon interaction \cite{Otsuka}. It is in general believed that shape coexistence can be interpreted microscopically in terms of particle-hole excitations \cite{Wood,Heyde} occurring across major shell and subshell closures \cite{Sorlin}, like the ones occurring at $Z=82$, 50, 40. However, the occurrence of shape coexistence in the region of the $N=90$ isotones, which are well-known examples of manifestation \cite{McCutchan} of the X(5) critical point symmetry \cite{IacX5}, related to the shape/phase transition \cite{Cejnar} between spherical and deformed nuclei, remains an open question (\cite{Garrett}, p. 104). 

In this Letter we suggest that the occurrence of shape coexistence in the $N=90$ region and in the similar regions at $N=70$, 40 can also be interpreted in terms of particle-hole excitations, which are however occurring across 3-dimensional isotropic harmonic oscillator (3D-HO) magic numbers. The main characteristic of these excitations is that they allow the involvement of orbitals coming down, because of the spin-orbit interaction \cite{Mayer}, from the next 3D-HO shell, which bear the opposite parity and are known to be of utmost importance for the onset of nuclear deformation \cite{FP,Casten,Cakirli,EPJASM,SDANCA21}. 

Standard covariant density functional theory is used for our calculations. In particular, the DDME2 functional of Ref. \cite{Lalazissis} is used within the code of Ref. \cite{Niksic}. The single particle energies are determined, labeled by Nilsson quantum numbers \cite{NR}, using the method described in Refs. \cite{Prassa,Karakatsanis1,Karakatsanis2}.

In Section 2 we present CDFT results for the Z=84, 52, 40 isotopes, in which particle-hole excitations across the Z=82, 50, 40 shell and subshell closures are known to exist \cite{Wood,Heyde}. In addition to demonstrating the applicability of the CDFT approach for the study of particle-hole excitations, the present results indicate that p-h excitations do not take place everywhere, but within specific regions around the neutron midshell. In Section 3 we present CDFT results for the N=92, 90, 60, 38 isotones, demonstrating that particle-hole excitations across the 3D-HO magic numbers 112, 70, 40 occur in nuclei exhibiting shape coexistence in these regions.

\section{Neutron induced shape coexistence} 

The series of isotopes close to the Z=82 magic number are known to provide textbook examples of shape coexistence \cite{Wood,Heyde,Garrett}. This is the most robust region of shape coexistence, thus we start from it and move to lower masses as we go on. 
Results for the Po (Z=84) isotopes are shown in Fig. 1. 4p-4h proton excitations are observed for N=98-108. The orbitals 11/2[505] (a), 1/2[400] (b) (normally lying below Z=82) are vacant, while the orbitals 1/2[541], 3/2[532] (c) (normally lying above Z=82) are occupied. Similar results for the Pb (Z=82) and Hg (Z=80) isotopes have been reported in \cite{EPJASC}. In the Pb isotopes, similar p-h excitations are seen again for N=98-108, while in the Hg isotopes they occur for a slightly wider region, N=96-110 \cite{EPJASC}.  

The series of isotopes close to the Z=50 magic number are also known to provide textbook examples of shape coexistence \cite{Wood,Heyde,Garrett}. 
Results for the Te (Z=52) isotopes, lying close to the Z=50 magic number,  are shown in Fig. 2. 2p-2h proton excitations are observed for N=64-68. The orbital 9/2[404] (a) (normally lying below Z=50) is vacant, while the orbital 3/2[422] (b) (normally lying above Z=50) is occupied. 

Results for the Zr (Z=40) isotopes, lying at the Z=40 subshell closure, are shown in Fig. 3. 6p-6h proton excitations are observed for N=38-40. The orbitals 1/2[301], 3/2[301], 5/2[303] (a) (normally lying below Z=40) are vacant, while the orbitals 1/2[440], 3/2[431], 5/2[422]  (b) (normally lying above Z=40) are occupied. 

In the cases mentioned above, it is clear that proton p-h excitations do not appear along the whole isotope chains, but are confined in regions near the neutron mid-shells, which are 104, 66, 39 respectively. This is in agreement with the interpretation that protons are elevated (thus creating proton p-h excitations) driven by the neutrons through the proton-neutron interaction. The proton-neutron interaction is expected to be maximized around the mid-shell, since in the beginning of the shell the available valence neutrons are few, while near the end of the shell the available valence neutron holes are few \cite{EPJAHW}. Counting of the valence neutrons from the nearest closed shell has been demonstrated to hold through the $N_pN_n$ scheme \cite{Casten}, for example. It is also one of the basic assumptions of the Interacting Boson Model \cite{IAbook}.   

The above approach corroborates the conclusion of Ref. \cite{EPJASC}, in which it is proved, using the proxy-SU(3) symmetry \cite{proxy1,proxy2}, that shape coexistence can occur only within certain islands of the nuclear chart, lying between 7-8, 17-20, 34-40, 59-70, 96-112, 146-168 nucleons. In the regions close to Z=82 and Z=50, which lie outside these limits, shape coexistence occurs because the neutrons lie within these limits, thus we call this case neutron induced shape coexistence. The Z=40 case will be further discussed below. 
 
\section{Proton induced shape coexistence} 

Shape coexistence is known to occur in the region with Z=64, N=90 \cite{Wood,Heyde}, but its microscopic origin remains an open question (\cite{Garrett}, p. 104).  We are going to attempt to provide a justification of the appearance of shape coexistence in this region in terms of particle-hole excitations, but we are going to consider p-h excitations of neutrons in this case, in contrast to the proton p-h excitations considered in Sec. 2. Results for the N=90 isotones are shown in Fig. 4. 2p-2h neutron excitations are observed for Z=60-64. The orbital 5/2[523] (a) (lying below the N=112 3D-HO magic number) is vacant, while the orbital 1/2[660] (b) (lying above the N=112 3D-HO magic number) is occupied. In addition, results for the N=92 isotones are shown in Fig. 5. 2p-2h neutron excitations are again observed for Z=60-64. The orbital 5/2[523] (a) (lying below the N=112 3D-HO magic number) is vacant, while the orbital 3/2[651] (b) (lying above the N=112 3D-HO magic number) is occupied.

In the cases mentioned above, it is clear that neutron p-h excitations do not appear along the whole isotone chains, but are confined in a region near to the proton mid-shell, which is 66. This is in agreement with the interpretation that neutrons are elevated (thus creating neutron p-h excitations) driven by the protons through the proton-neutron interaction. The picture is completely analogous to the one described in the previous section, in which protons get elevated by the neutrons. However, since in the nuclei under consideration in most cases $N>Z$ is valid, it is qualitatively expected that neutron elevation by (relatively fewer) protons will be weaker than proton elevation by (relatively more) neutrons. We call the present case proton induced shape coexistence.

The Z=64, N=90 region, considered above, is known to be very similar to the region with Z=40, N=60 (\cite{Garrett}, p. 79). Results for the N=60 isotones are shown in Fig. 6. 4p-4h proton excitations are observed for Z=40-42, near the Z=39 midshell. The orbitals 1/2[411], 5/2[413] (a) (lying below the N=70 3D-HO magic number) are vacant, while the orbitals 1/2[550], 3/2[541]  (b) (lying above the N=70 3D-HO magic number) are occupied.

A similar picture is seen in the case of the N=38 isotones, shown in Fig. 7. 4p-4h proton excitations are observed for Z=40. The orbitals 5/2[303], 3/2[301] (a) (lying below the N=40 3D-HO magic number) are vacant, while the orbitals 1/2[440], 3/2[431]  (b) (lying above the N=40 3D-HO magic number) are occupied. The present case is related to the third case of  the previous section. Both fall within the regions of 34-40 protons and 34-40 neutrons. In these cases the observed shape coexistence is both proton induced and neutron induced. It is clear in this case that one should take into account the fact that protons and neutrons occupy the same shells, therefore the Wigner SU(4) supermultiplet may play an enhanced role \cite{KS}.  

\section{Discussion}

Several comments are in place.

a) In the cases of neutron induced shape coexistence, proton p-h excitations across major shell closures (Z=82, 50) or well known subshell closures (Z=40) occur. In contrast, in the proton induced shape coexistence, neutron p-h excitations occur across 3D-HO magic numbers (40, 70, 112). One could wonder what is the physical meaning of such p-h excitations, since it is well known that beyond the sd shell (20 protons, 20 neutrons), the 3D-HO magic numbers 40, 70, 112, 168, \dots are destroyed by the spin-orbit interaction, giving rise to the shell model magic numbers 28, 50, 82, 126, 184, \dots \cite{Mayer}. The answer lies in the fact that the orbitals lying in the 40-50, 70-82, 112-126 intervals, are respectively the 1g9/2, 1h11/2, 1i13/2 intruder orbitals pushed by the spin-orbit interaction into the 3D-HO shell below their original one. The intruder orbitals posses parity opposite to the parity of their new partners in the shell they enter, and are known to play a leading role in the creation of nuclear deformation \cite{FP,Casten,Cakirli,EPJASM,SDANCA21}. Thus, for example, p-h excitations across N=112 mean that some of the negative parity orbitals residing in the 82-126 shell (which are the 3p1/2, 3p3/2, 2f5/2, 2f7/2, 1h9/2 orbitals) get empty, while some of the positive parity orbitals of 1i13/2 get occupied. It should be emphasized that, as known from the standard Nilsson diagrams \cite{NR}, the intruder 1i13/2 orbital does not appear at the top of the 82-126 neutron shell, but it is rather located in the middle of it, while several of its suborbitals rapidly fall below midshell with increasing deformation.  Therefore it makes sense to talk about p-h excitations across the N=112 magic number of the 3D-HO in the N=90 region. 

b) Similarly, p-h excitations across N=70 mean that some of the positive parity orbitals residing in the 50-82 shell (which are the 3s1/2, 2d3/2, 2d5/2, 1g7/2 orbitals) get empty, while some of the negative parity orbitals of 1h11/2 get occupied. As seen in the standard Nilsson diagrams \cite{NR}, the intruder orbital 1h11/2 lies at the upper part of the 50-82 neutron shell for zero deformation, but several of its suborbitals fall rapidly to midshell with increasing deformation.  Therefore it makes sense to talk about p-h excitations across the N=70 magic number of the 3D-HO in the N=60 region.

c) In other words, p-h excitations across a 3D-HO magic number mean that the intruder orbitals start being occupied, thus triggering the enhancement of proton-neutron interaction and the onset of deformation
 \cite{FP,Casten,Cakirli}.  This is clearly seen in the N=90 isotones with Z=60-64, which are $^{150}$Nd, $^{152}$Sm, and $^{154}$Gd, the well known best manifestations \cite{McCutchan} of the X(5) critical point symmetry \cite{IacX5} of the shape/phase transition \cite{Cejnar} from spherical to deformed shapes. 

d) The previous point clarifies and corroborates the relation between shape coexistence and shape/phase transitions, considered in detail in the Z=40, N=60 region in \cite{Ramos}, within the framework of the interacting boson model with configuration mixing \cite{IAbook}.  

e) All cases of shape coexistence discussed above fall within the islands of shape coexistence predicted through the proxy-SU(3) symmetry \cite{EPJASC}. 

In conclusion,  a microscopic explanation of the nature of shape coexistence in the N=90, Z=64 region is suggested. Shape coexistence in the N=90 region is caused by the protons, which create neutron p-h excitations across the N=112 3D-HO magic number, signaling the start of the occupation of the 1i13/2 intruder orbital, which triggers stronger proton-neutron interaction, causing the onset of the deformation and resulting in the shape/phase transition from  spherical to deformed nuclei described by the X(5) critical point symmetry. A similar picture holds in the N=60, Z=40 region, in which  p-h excitations across the N=70 3D-HO magic number occur, signaling the start of the occupation of the 1h11/2 intruder orbital. 

The results of a more exhaustive search for islands of shape coexistence across the nuclear chart, using the same method, is deferred to a longer publication. Preliminary results indicate that neutron induced shape coexistence, based on proton p-h excitations, is also found also in the Z=50 region, while proton induced shape coexistence, based on neutron p-h excitations, is also found also in the N=60 region around Z=40. 

Support by the Tenure Track Pilot Programme of the Croatian Science Foundation and the Ecole Polytechnique F\'{e}d\'{e}rale de Lausanne, the Project TTP-2018-07-3554 Exotic Nuclear Structure and Dynamics with funds of the Croatian-Swiss Research Programme, 
as well as by the Bulgarian National Science Fund (BNSF) under Contract No. KP-06-N48/1  is gratefully acknowledged.



\begin{figure*}[htb]

\includegraphics[width=75mm]{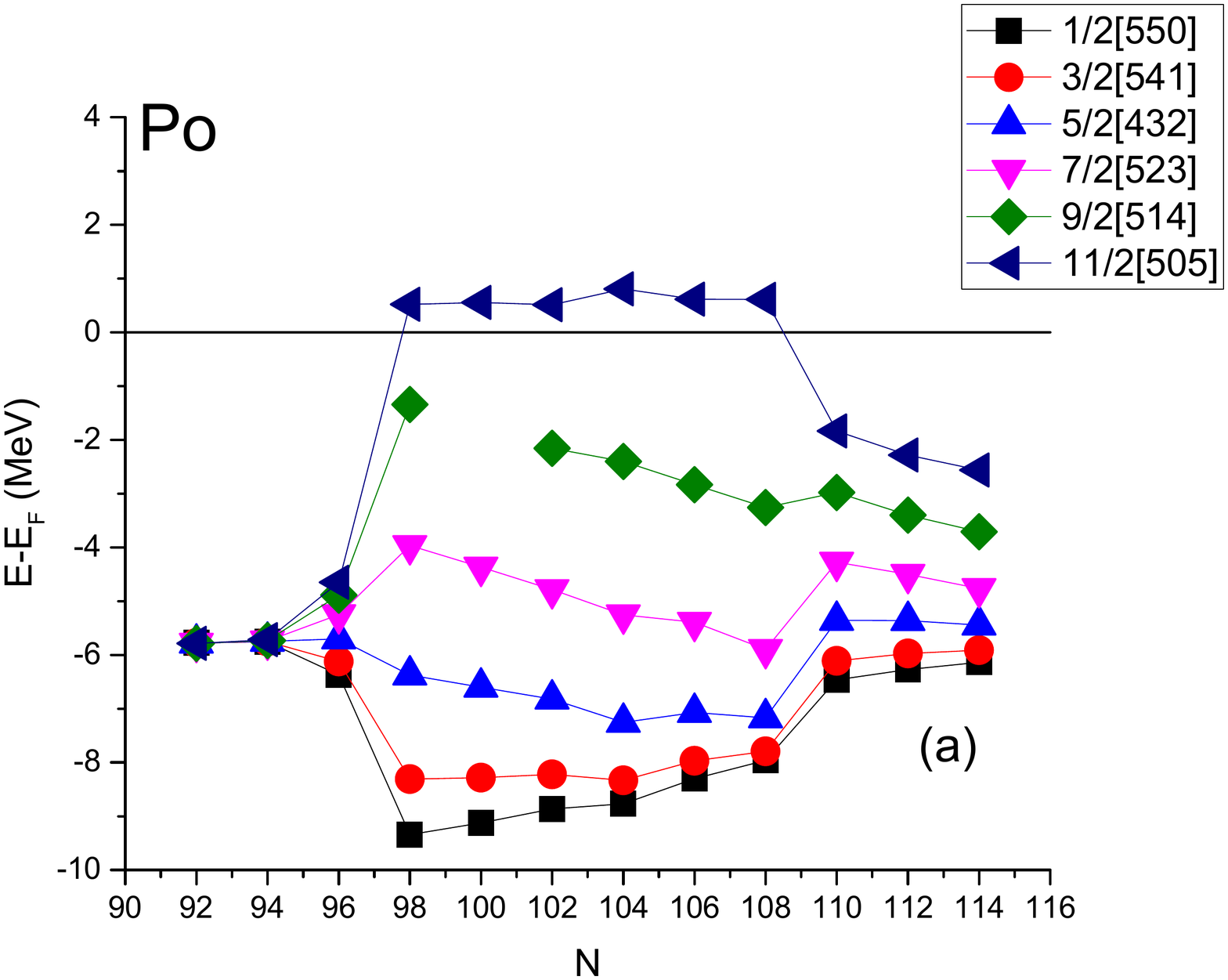}
\includegraphics[width=75mm]{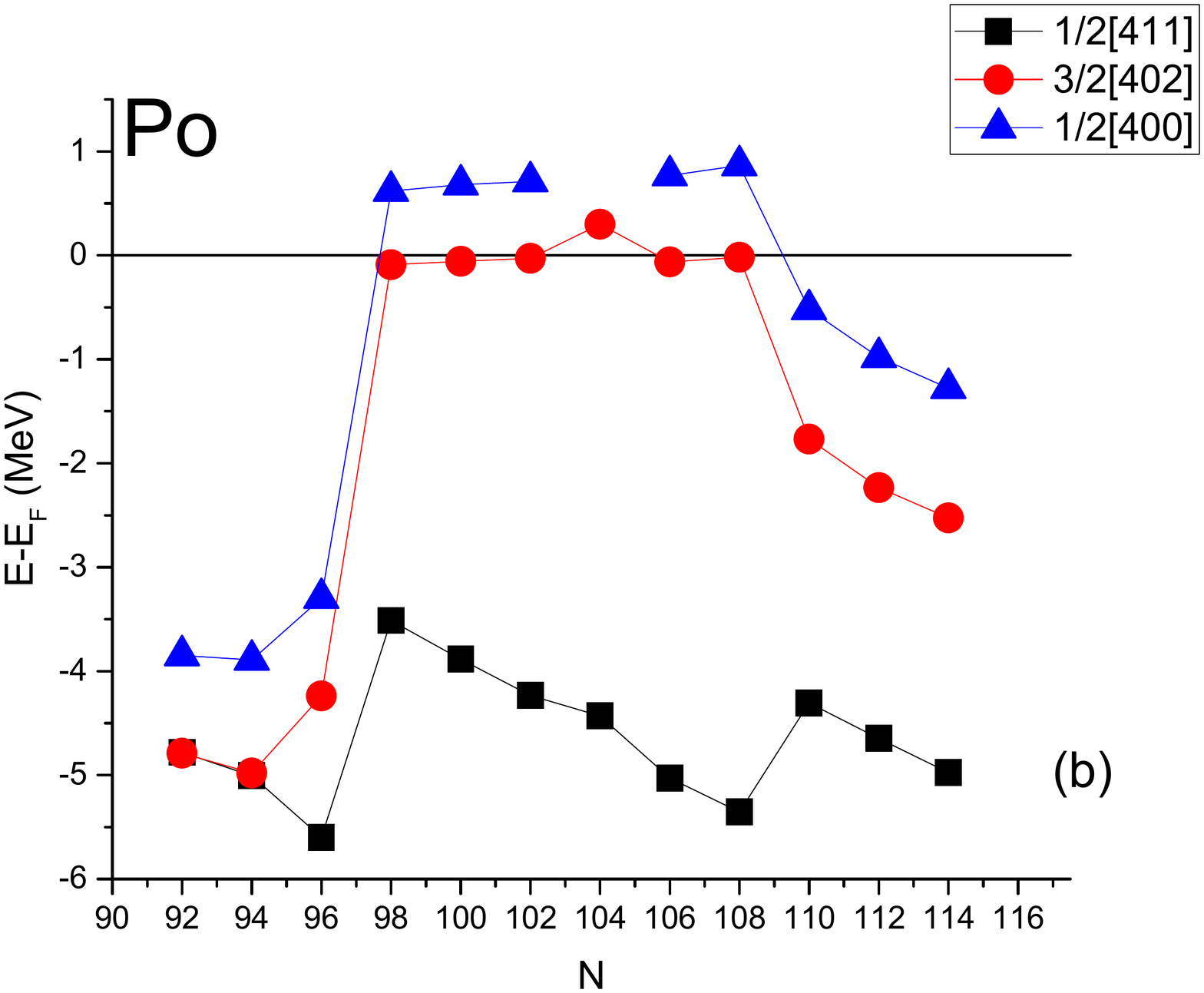}
\includegraphics[width=75mm]{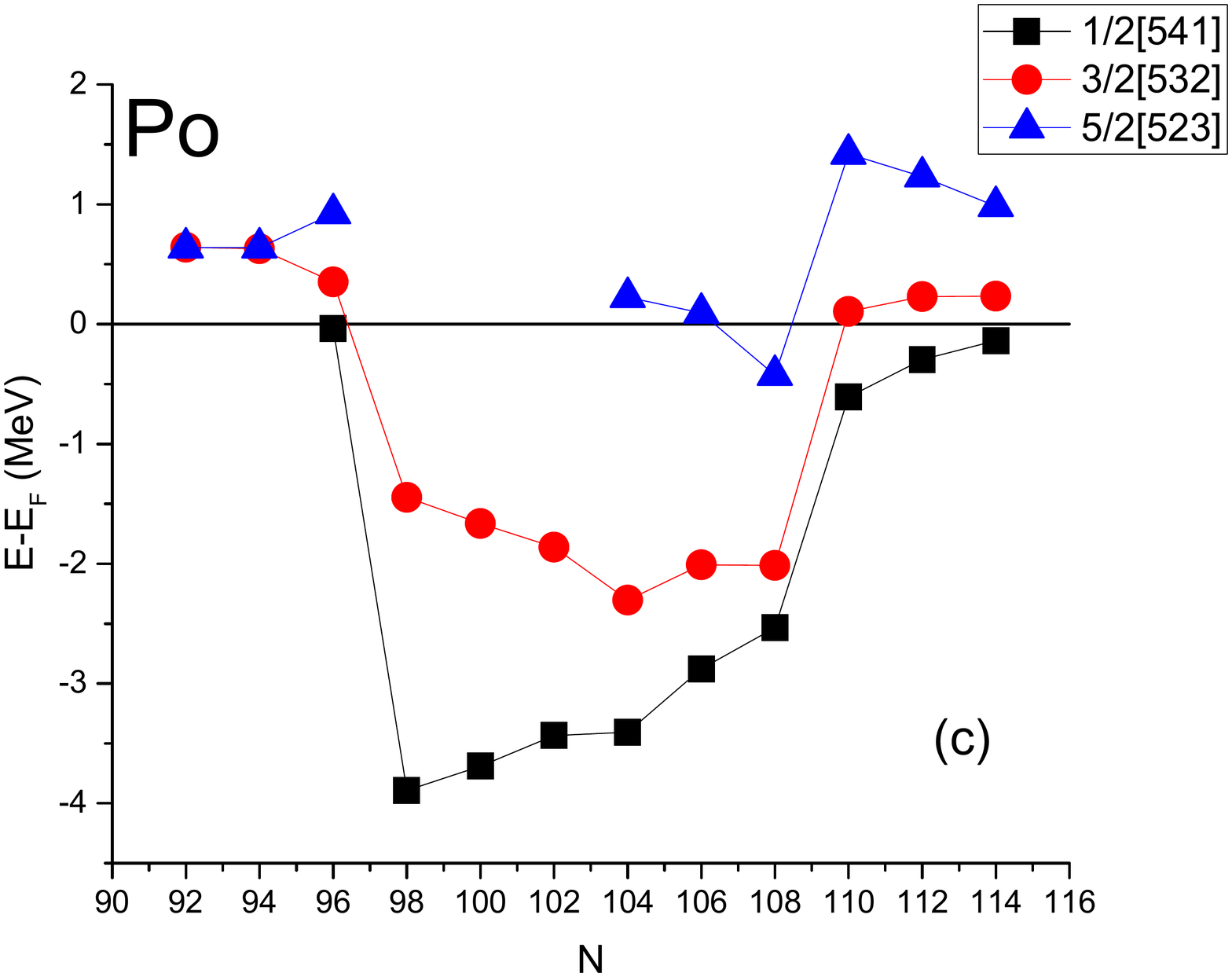}

\caption{Energies (in MeV) of proton single particle orbitals relative to the Fermi energy obtained by CDFT for Po (Z=84) isotopes. 4p-4h proton excitations observed for N=98-108. The orbitals 11/2[505] (a), 1/2[400] (b) (normally lying below Z=82) are vacant, while the orbitals 1/2[541], 3/2[532] (c) (normally lying above Z=82) are occupied. See Section 2 for further discussion. 
} 

\end{figure*}


\begin{figure*}[htb]

{\includegraphics[width=75mm]{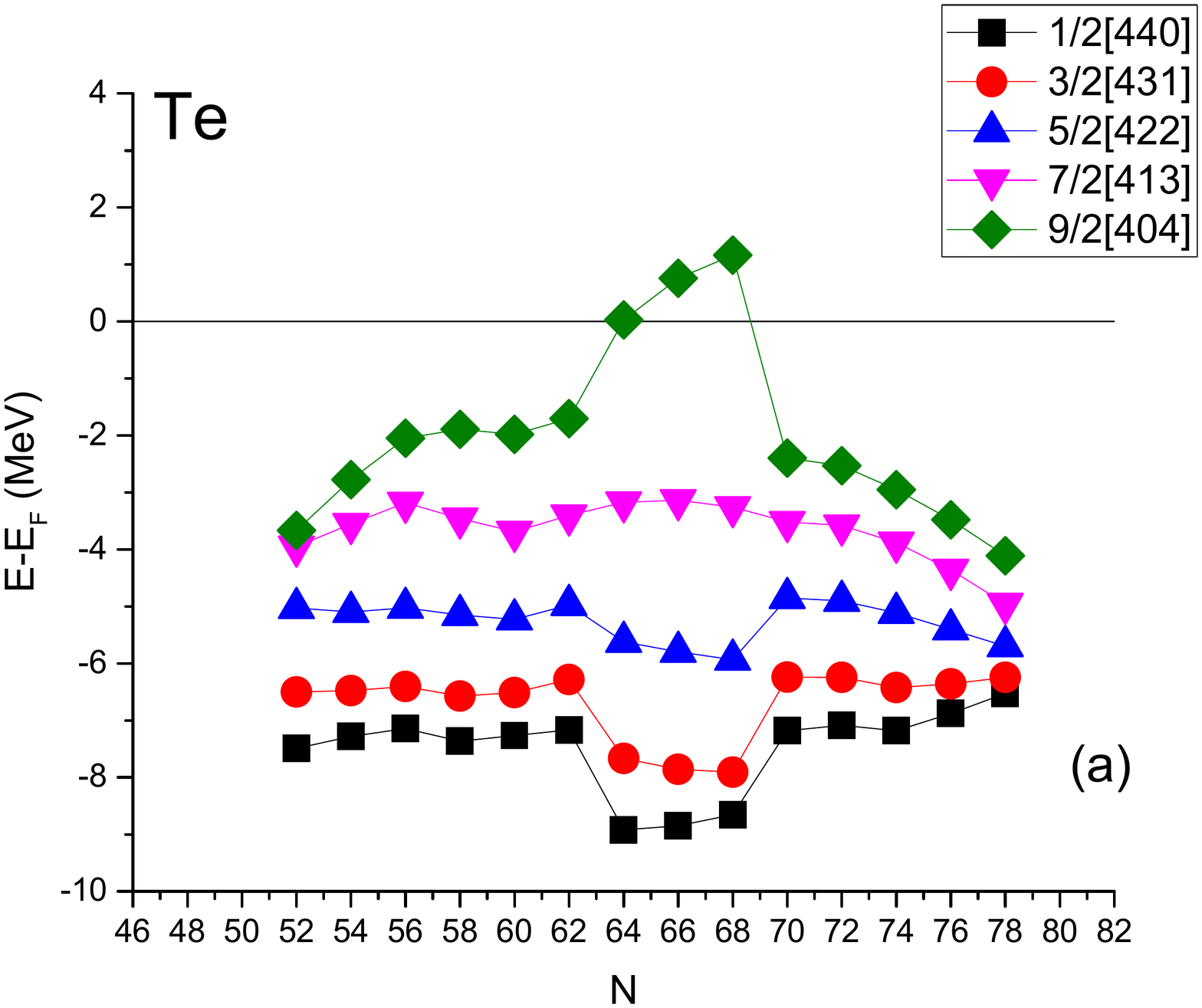}\hspace{5mm}
\includegraphics[width=75mm]{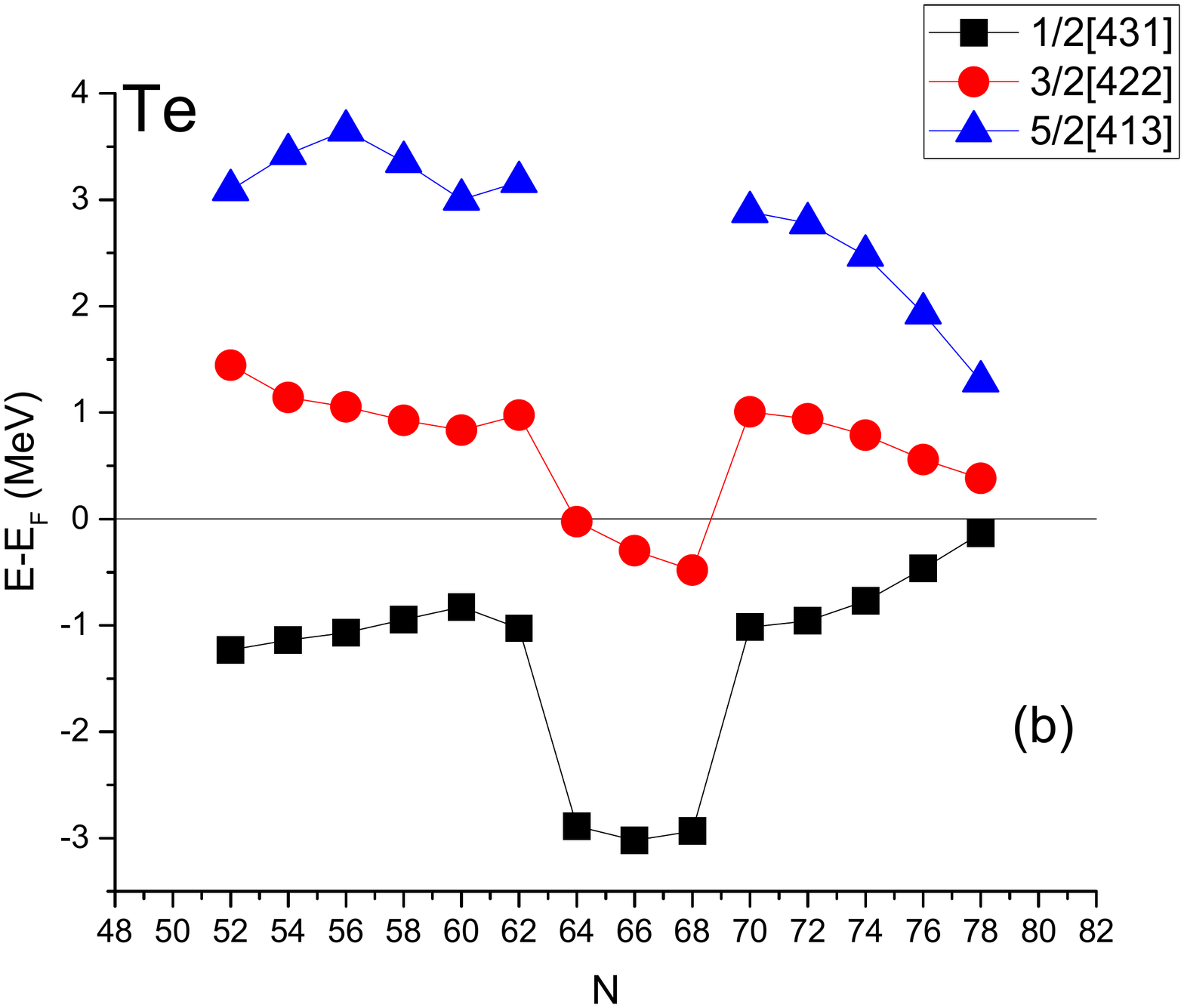}}

\caption{Energies (in MeV) of proton single particle orbitals relative to the Fermi energy obtained by CDFT for Te (Z=52) isotopes. 2p-2h proton excitations observed for N=64-68. The orbital 9/2[404] (a) (normally lying below Z=50) is vacant, while the orbital 3/2[422] (b) (normally lying above Z=50) is occupied. See Section 2 for further discussion. 
} 

\end{figure*}


\begin{figure*}[htb]

{\includegraphics[width=75mm]{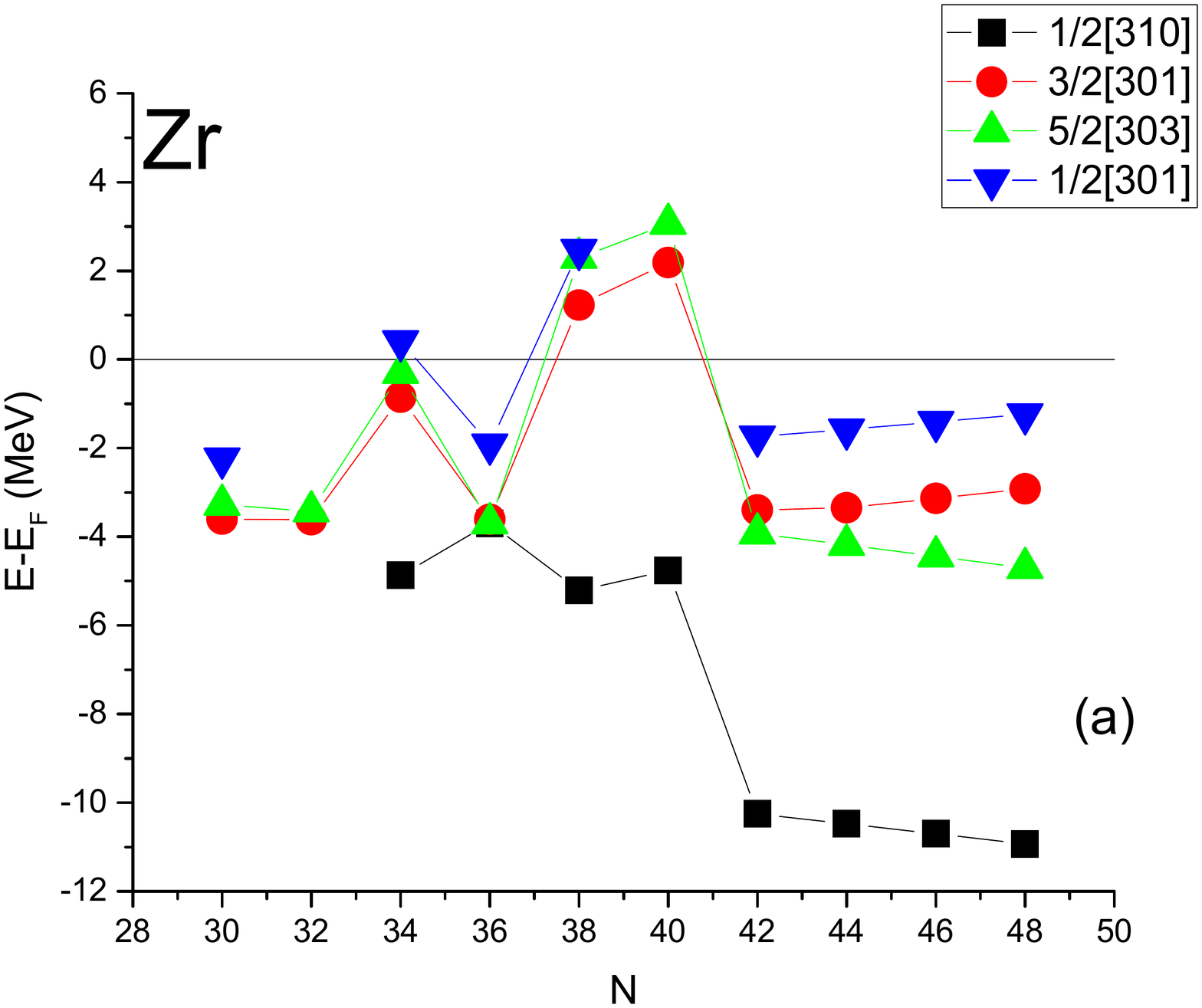}\hspace{5mm}
\includegraphics[width=75mm]{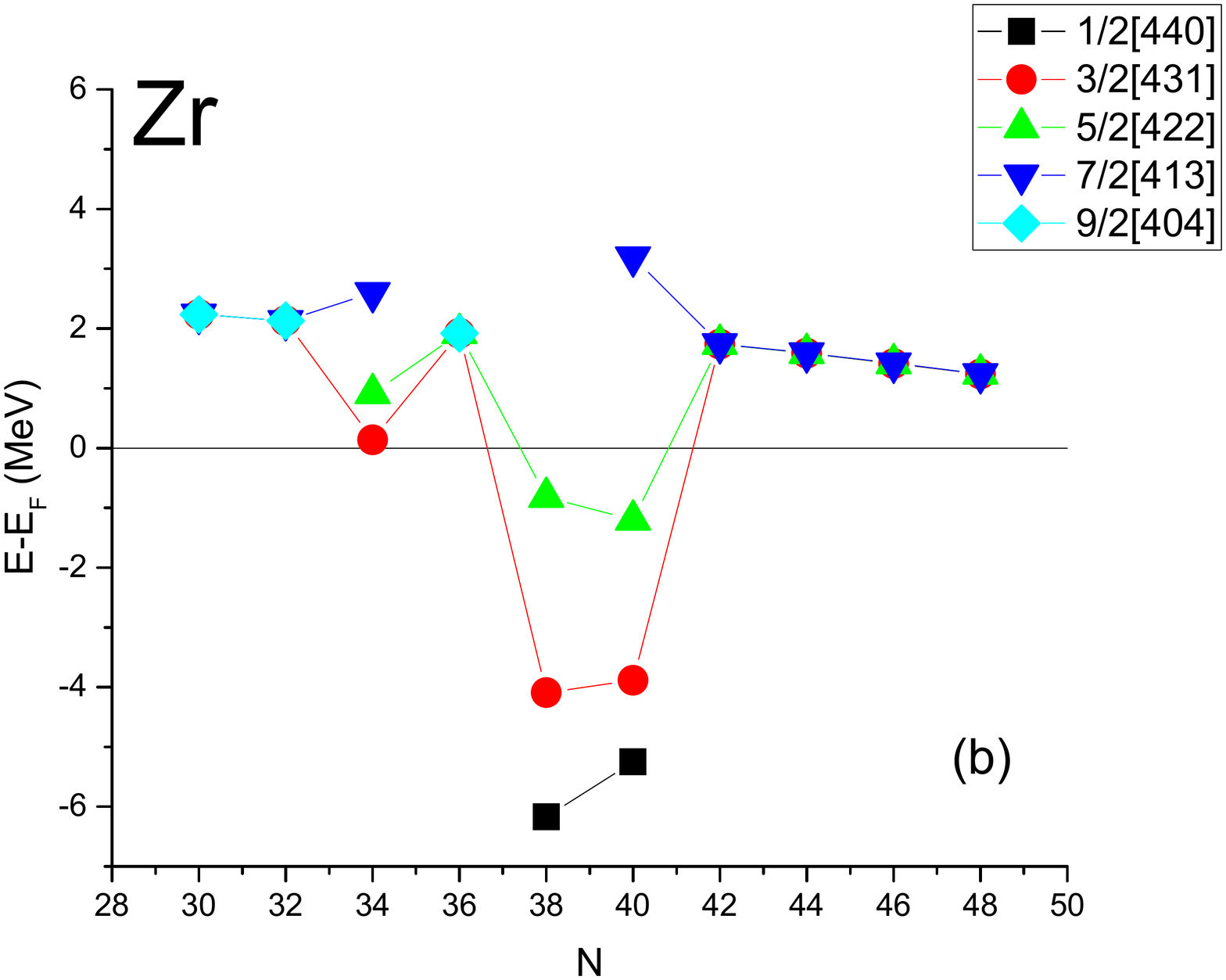}}

\caption{Energies (in MeV) of proton single particle orbitals relative to the Fermi energy obtained by CDFT for Zr (Z=40) isotopes. 6p-6h proton excitations observed for N=38-40. The orbitals 1/2[301], 3/2[301], 5/2[303] (a) (normally lying below Z=40) are vacant, while the orbitals 1/2[440], 3/2[431], 5/2[422]  (b) (normally lying above Z=40) are occupied. See Section 2 for further discussion. 
} 

\end{figure*}


\begin{figure*}[htb]

{\includegraphics[width=75mm]{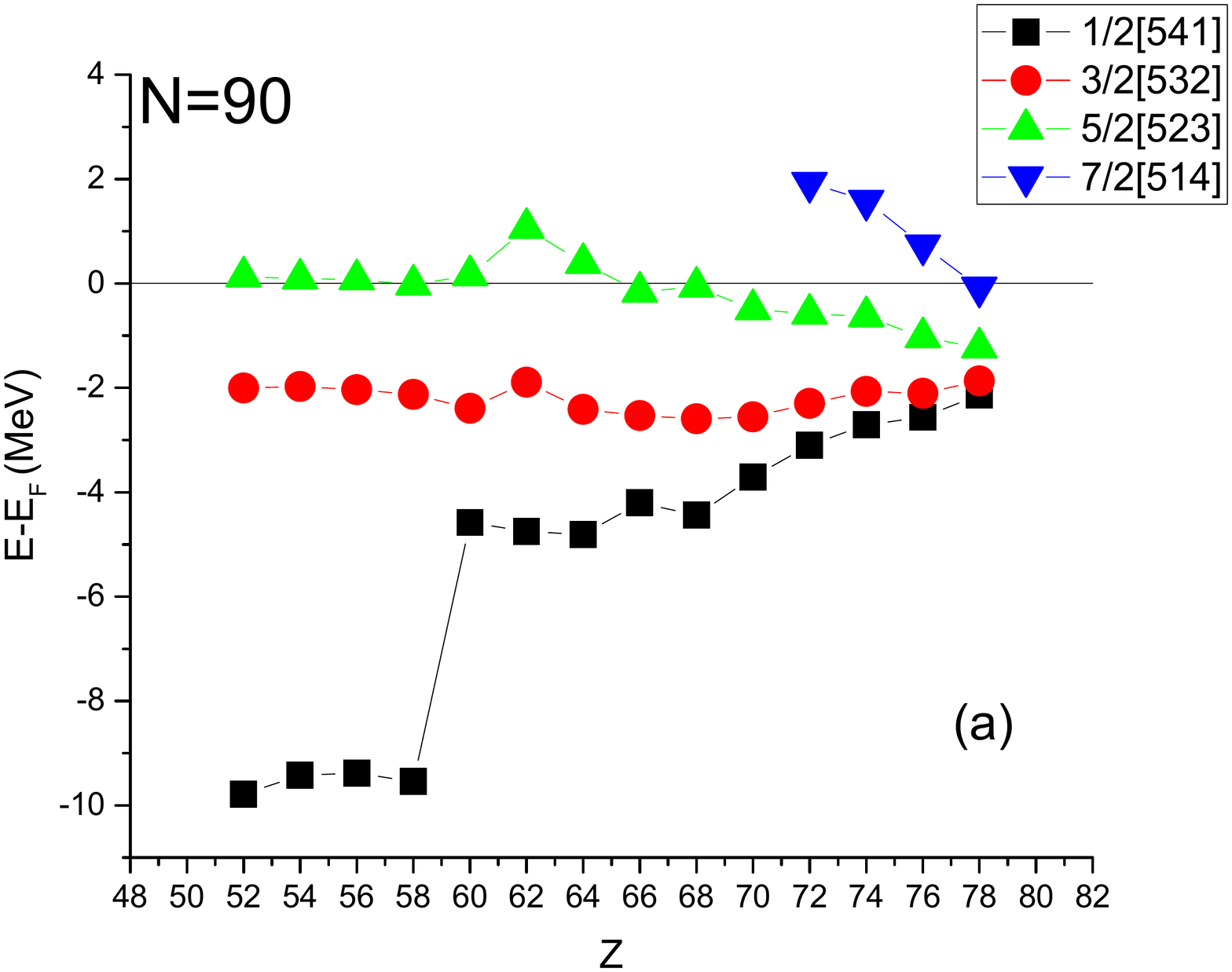}\hspace{5mm}
\includegraphics[width=75mm]{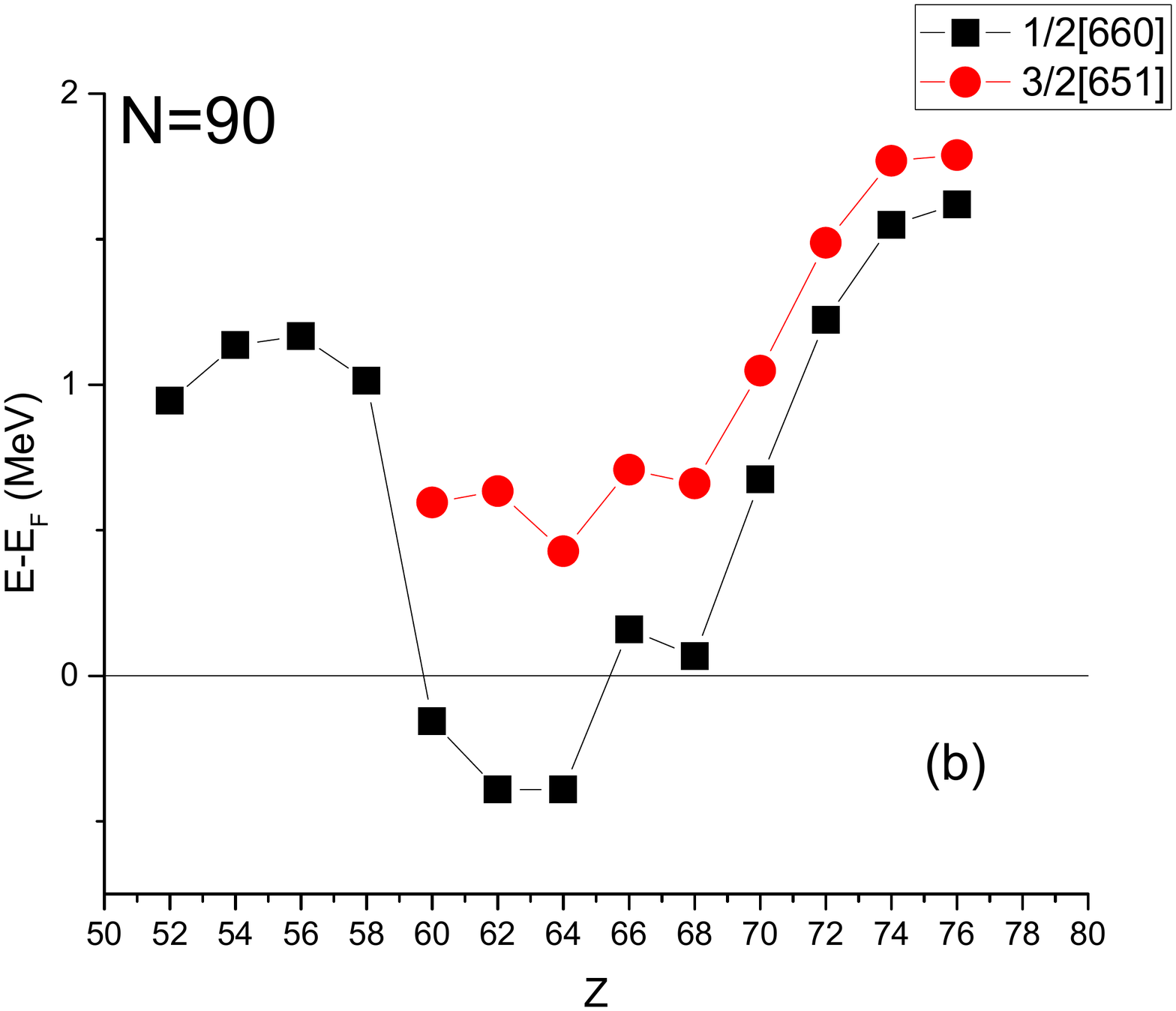}}

\caption{Energies (in MeV) of neutron single particle orbitals relative to the Fermi energy obtained by CDFT for N=90 isotones. 2p-2h neutron excitations observed for Z=60-64. The orbital 5/2[523] (a) (lying below the N=112 3D-HO magic number) is vacant, while the orbital 1/2[660] (b) (lying above the N=112 3D-HO magic number) is occupied. See Section 3 for further discussion. 
} 

\end{figure*}


\begin{figure*}[htb]

{\includegraphics[width=75mm]{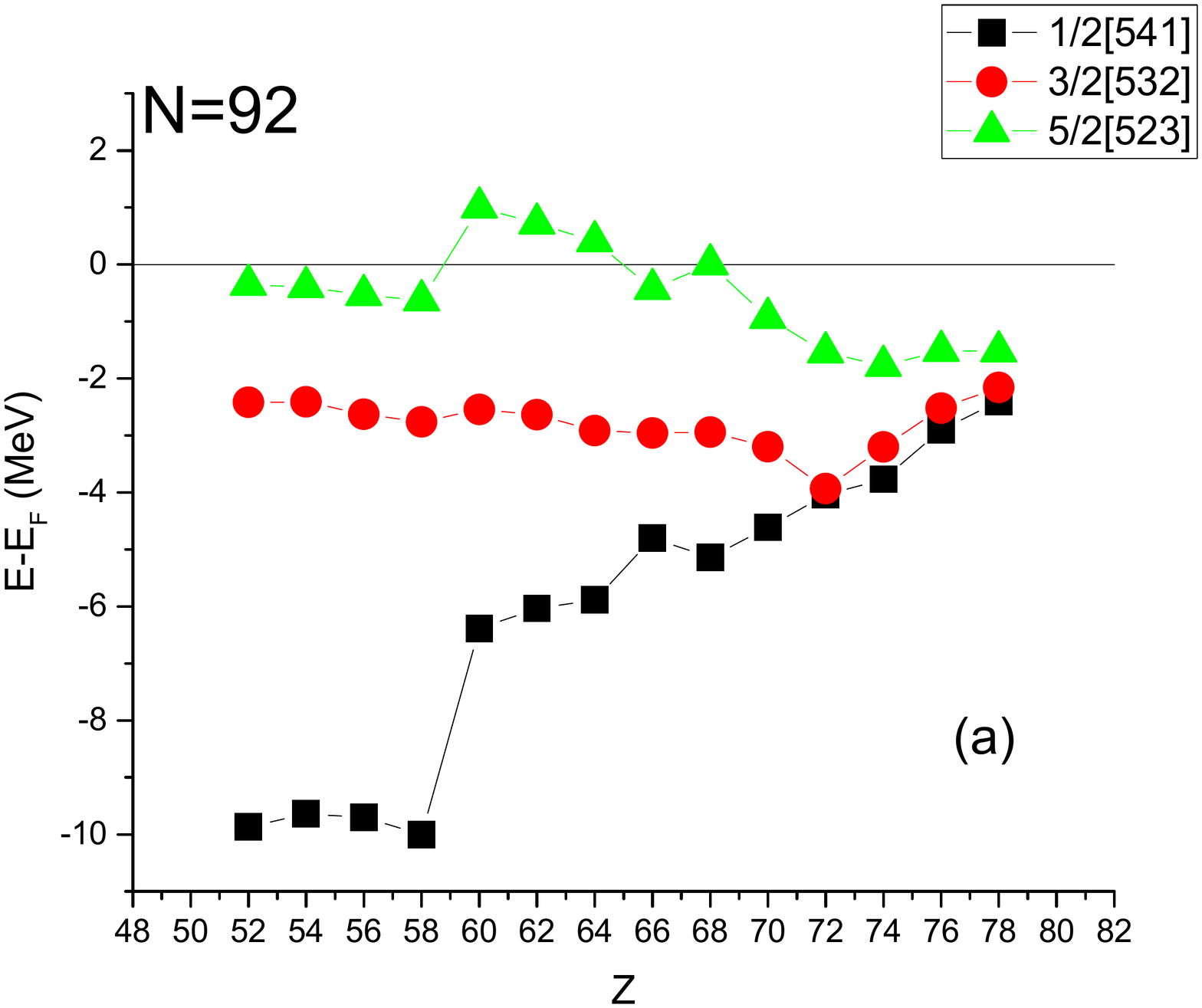}\hspace{5mm}
\includegraphics[width=75mm]{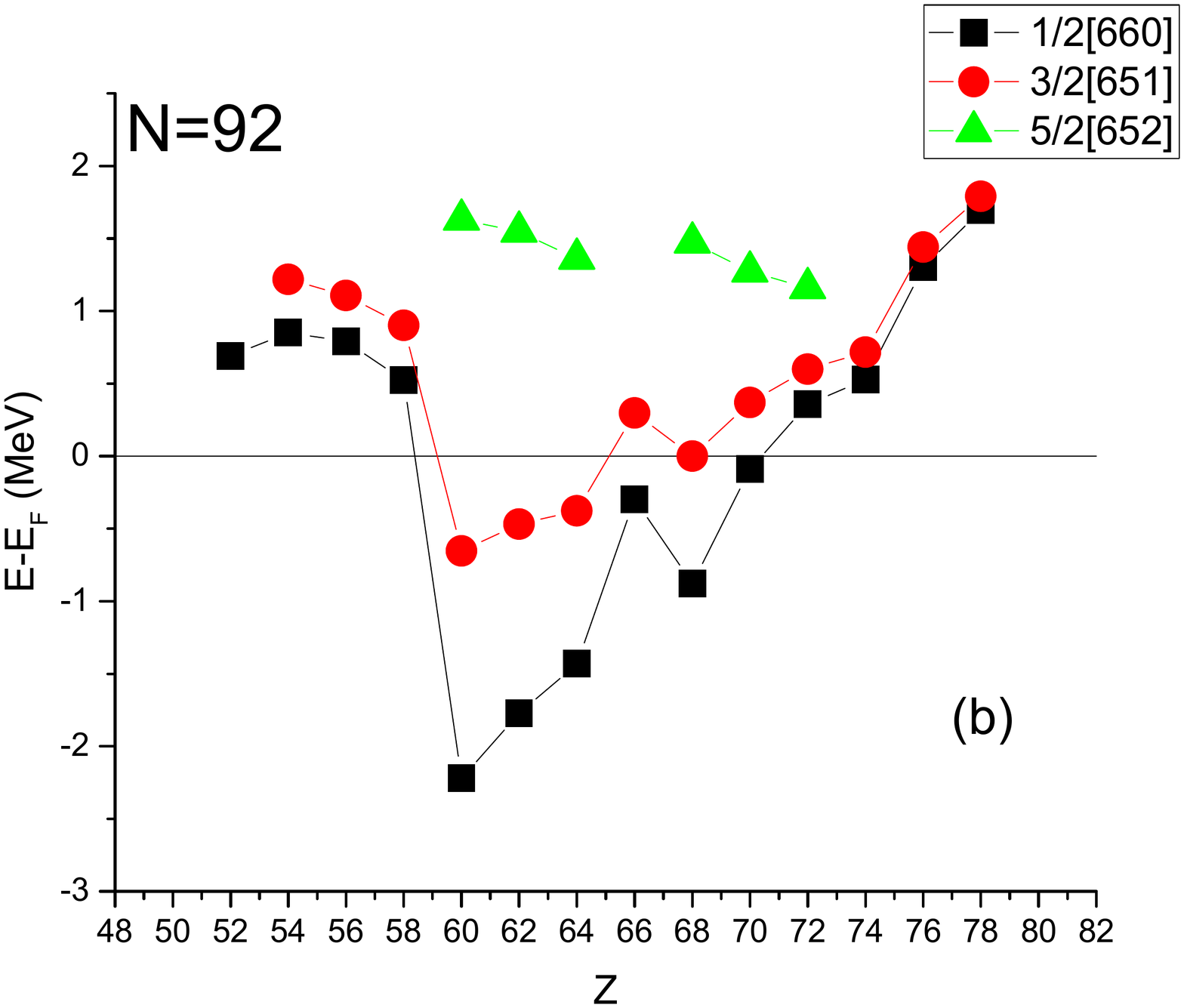}}

\caption{Energies (in MeV) of neutron single particle orbitals relative to the Fermi energy obtained by CDFT for N=92 isotones. 2p-2h neutron excitations observed for Z=60-64. The orbital 5/2[523] (a) (lying below the N=112 3D-HO magic number) is vacant, while the orbital 3/2[651] (b) (lying above the N=112 3D-HO magic number) is occupied. See Section 3 for further discussion. 
} 

\end{figure*}


\begin{figure*}[htb]

{\includegraphics[width=75mm]{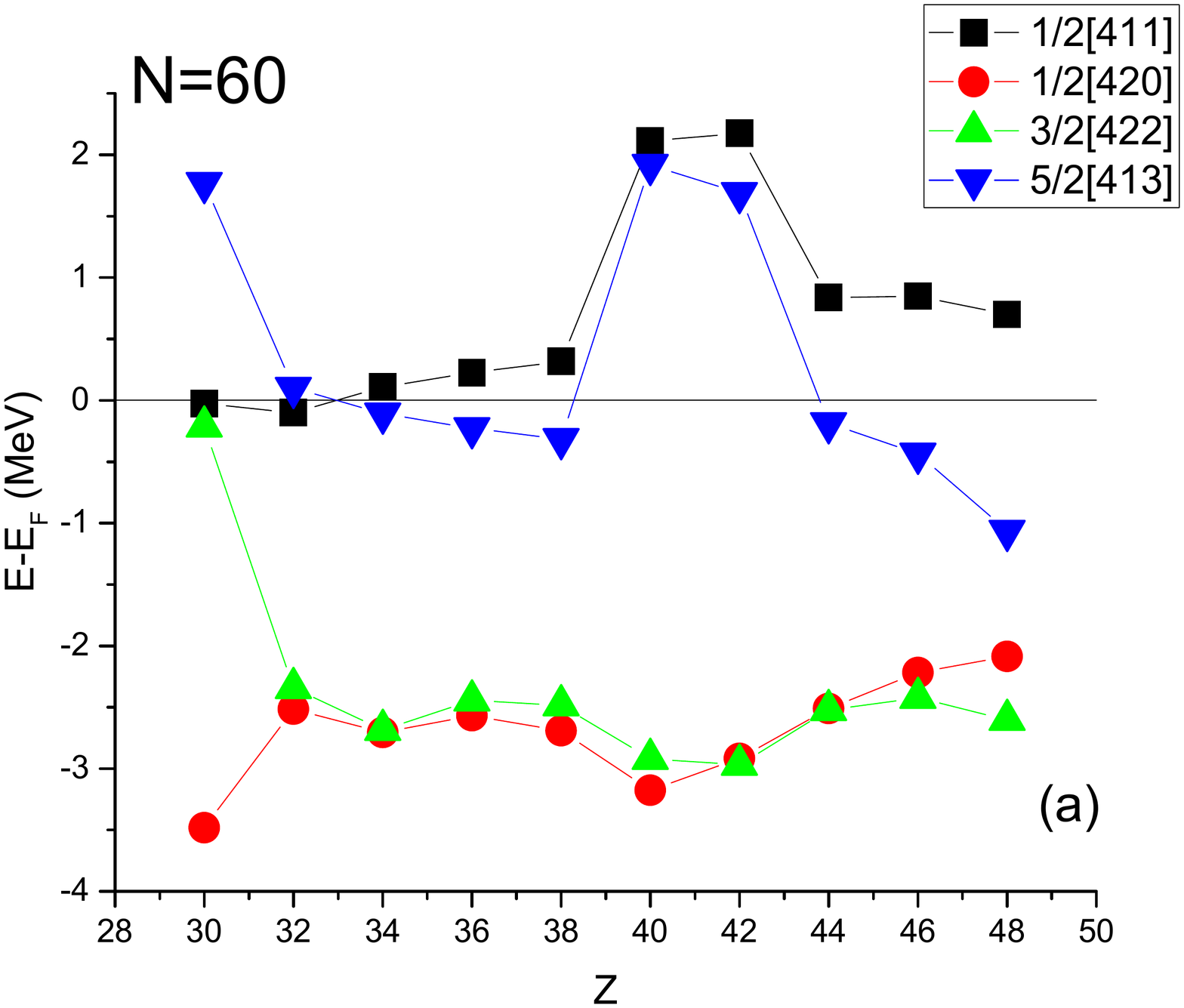}\hspace{5mm}
\includegraphics[width=75mm]{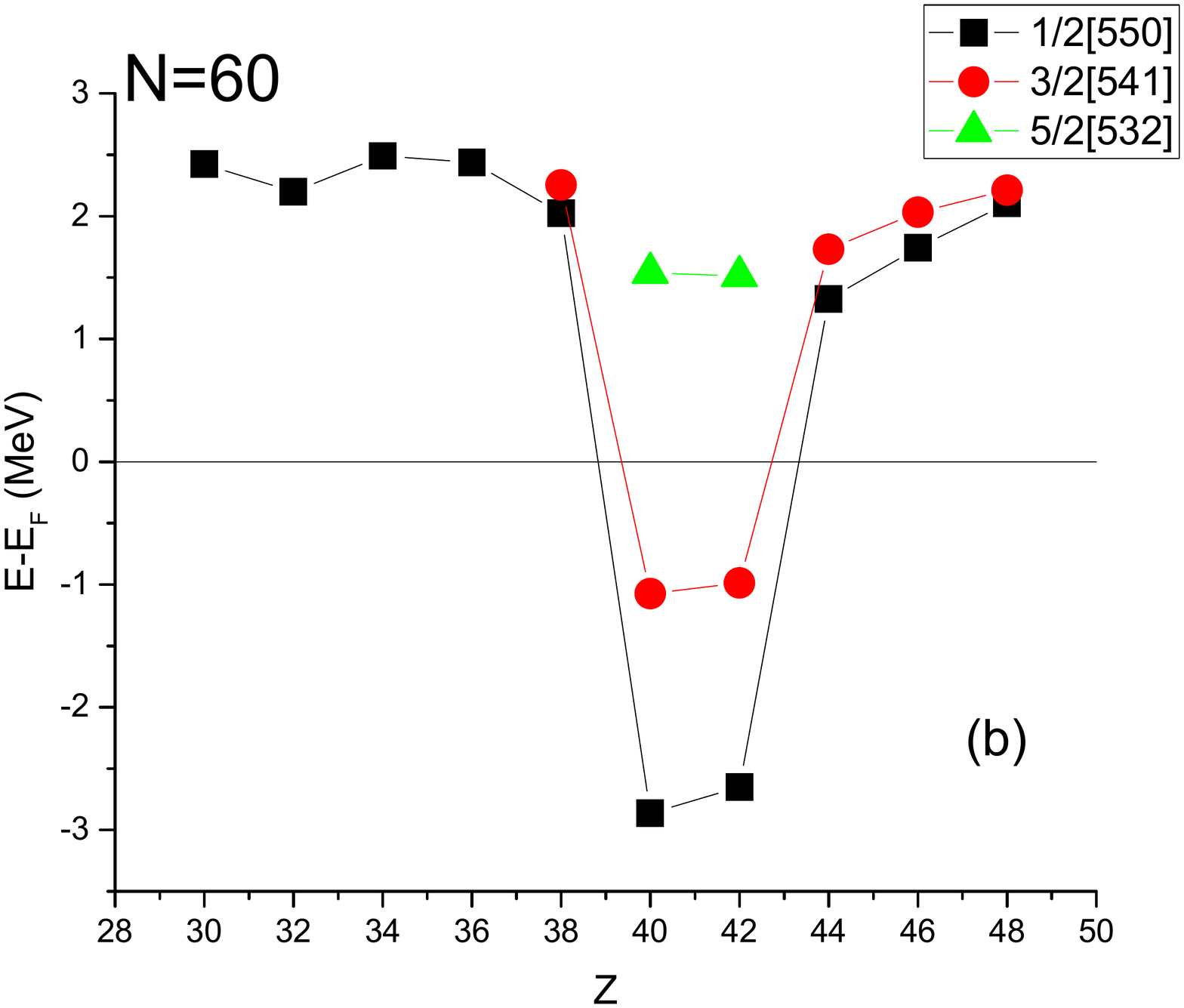}}

\caption{Energies (in MeV) of proton single particle orbitals relative to the Fermi energy obtained by CDFT for N=60 isotones. 4p-4h proton excitations observed for Z=40-42. The orbitals 1/2[411], 5/2[413] (a) (lying below the N=70 3D-HO magic number) are vacant, while the orbitals 1/2[550], 3/2[541]  (b) (lying above the N=70 3D-HO magic number) are occupied. See Section 3 for further discussion. 
} 

\end{figure*}


\begin{figure*}[htb]

{\includegraphics[width=75mm]{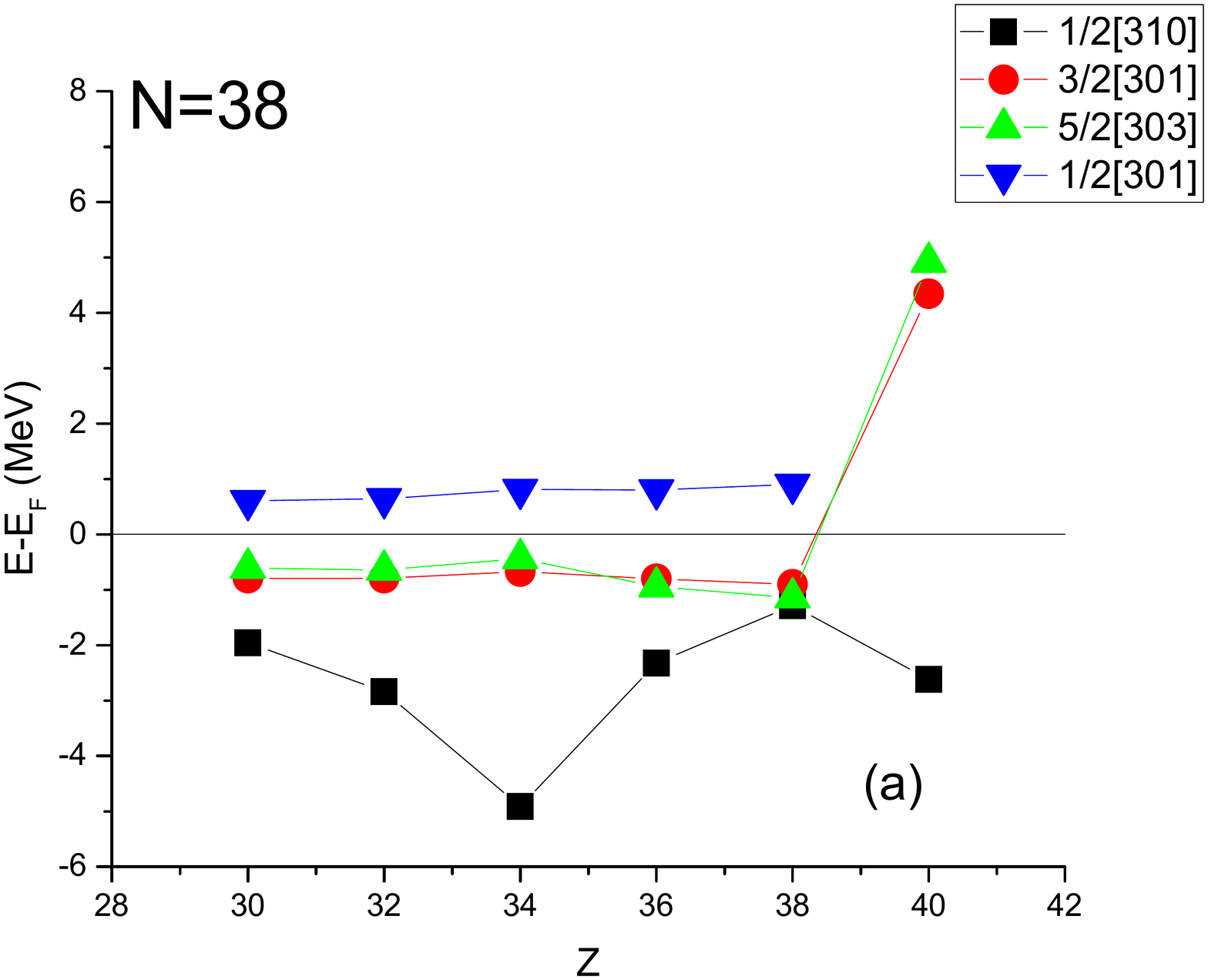}\hspace{5mm}
\includegraphics[width=75mm]{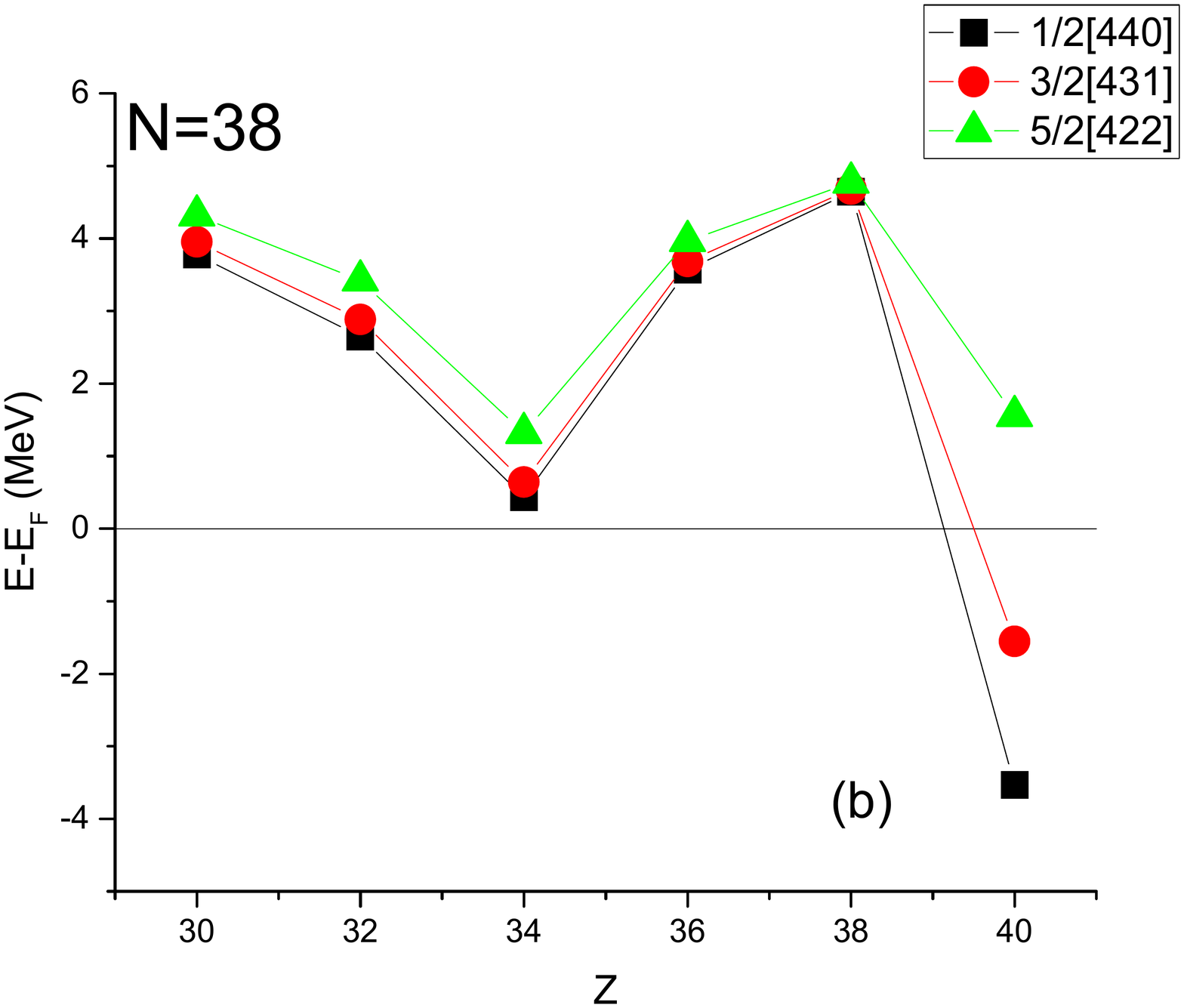}}

\caption{Energies (in MeV) of proton single particle orbitals relative to the Fermi energy obtained by CDFT for N=38 isotones. 4p-4h proton excitations observed for Z=40. The orbitals 5/2[303], 3/2[301] (a) (lying below the N=40 3D-HO magic number) are vacant, while the orbitals 1/2[440], 3/2[431]  (b) (lying above the N=40 3D-HO magic number) are occupied. See Section 3 for further discussion. 
} 

\end{figure*}

\end{document}